\documentclass[twocolumn, 10pt, conference]{IEEEtran}
\usepackage{color}
\usepackage{setspace}
\usepackage[cmex10]{amsmath}
\usepackage{psfrag}
\usepackage{subfigure}
\usepackage{hyperref}
\usepackage{multirow}
\usepackage{amsmath}
\usepackage{multicol}
\usepackage{setspace}
\usepackage{amsmath}
\usepackage{subfigure} 
\usepackage{multicol}
\usepackage{algorithm}
\usepackage{epsfig}
\usepackage{cite}
\usepackage{graphicx} 
\usepackage{geometry}
\usepackage{color}
\usepackage[space]{grffile}
\usepackage[english]{babel}
\usepackage[utf8]{inputenc}
\usepackage[noend]{algpseudocode}
\begin{document}
\renewcommand{\algorithmicrequire}{\textbf{Input:}}
\renewcommand{\algorithmicensure}{\textbf{Output:}}
\date{}
\title{{Modeling the Evolution of Line-of-Sight Blockage for V2V Channels}}
\author{
  		Mate Boban, Xitao Gong, and Wen Xu\\
       Huawei Technologies Duesseldorf GmbH,
       European Research Center \\
       Riesstr. 25c, 80992, Munich, Germany\\
	Email: \{mate.boban, xitao.gong, wen.dr.xu\}@huawei.com \\
}

\maketitle

\begin{abstract}

We investigate the evolution of line of sight (LOS) blockage over both time and space for vehicle-to-vehicle (V2V) channels. 
Using realistic vehicular mobility and building and foliage locations from maps, we first perform 
LOS blockage analysis to extract LOS 
probabilities in real cities and on highways 
for varying vehicular densities. 
Next, to 
model the time evolution of LOS blockage for V2V links, we employ a three-state discrete-time Markov chain comprised of the following states: i) LOS; ii) non-LOS due to static objects (e.g., buildings, trees, etc.); and iii) non-LOS due to mobile objects (vehicles). 
We obtain state transition probabilities based on the evolution of LOS blockage. 
Finally, we perform curve fitting and obtain a set of distance-dependent equations for both LOS and transition probabilities. These equations can be used to generate time-evolved 
V2V channel realizations for representative urban and highway environments. %
Our results can be used to perform highly efficient and accurate 
simulations without the need to employ complex geometry-based models for link evolution. 

\end{abstract}

\section{Introduction} \label{sec:Intro}

In contrast to existing cellular networks that focus primarily on increasing the mobile data rates, fifth-generation (5G) network is expected to efficiently support the so-called vertical industries (e.g., industrial, eHealth, and automotive vertical, among other). 
Supporting automotive vertical, in the form of vehicle-to-anything (V2X) communication, is seen as one of the most challenging tasks for 5G networks, 
particularly in terms of end-to-end latency and reliability~\cite{osseiran2014scenarios}.

One of the key distinguishing features of V2X communication is the high mobility, possibly on both sides of the link (as in the case of V2V communication). Another salient aspect of V2X communication is that it is 
often related to safety, either directly (e.g., emergency braking, intersection collision avoidance application~\cite{etsi14}, etc.) or indirectly (e.g., platooning, lane-change maneuvers, etc.). 
To ensure that V2X communication systems can support the application requirements 
efficiently, a  key initial step is realistically defining the channel characteristics for different environments (e.g., urban, highway, rural) and V2X communication types (e.g., V2V, V2I, V2P). Given the V2X application requirements, one of the most relevant aspects of channel modeling is the time evolution of V2V links and the related concept of spatial consistency. Time and space evolution of LOS blockage 
refers to time-consistent realization of LOS blockage for V2V channels, 
based on the location of the transmitter (Tx), the receiver (Rx), and the composition of their surroundings. 
Consistent LOS blockage realization is important in
order to assign the appropriate path loss, shadowing, small-scale,
and large-scale parameters over time and space.
This implies that there should be a continuity in terms of vehicle locations over time (i.e., requiring a continuous vehicle movement) and in terms of scatterer distribution around the vehicles. 

In the realm of the 3GPP channel modeling (e.g., SCME model~\cite{baum2005interim}), which resort to independent ``drops'' of devices in space and time, time evolution and spatial consistency have not been taken into consideration. On the other hand, time evolution and spatial consistency are inherently supported by geometry-based deterministic models (e.g.,~\cite{maurer2004new},~\cite{boban14TVT}). However, for 
performing efficient simulations it is also beneficial to have a model that can generate consistent channel realizations for a chosen environment without the need for complex geometric modeling using location-specific map information. While V2X channel measurements and modeling have attracted considerable attention in recent years~\cite{viriyasitavat2015vehicular}, a comprehensive geometry-based stochastic model for time and space evolution of LOS blockage 
is currently not available for V2V channels.

In this paper, we attempt to fill this gap by performing a comprehensive study of 
time evolution of V2V links in urban and highway environments. We employ a Markov chain comprised of three states: i) LOS; ii) NLOSb -- non-LOS due to static objects (e.g., buildings, trees, etc.); and iii) NLOSv -- non-LOS due to mobile objects (vehicles), since these three states were shown to have distinct path loss and shadowing parameters~\cite{abbas2012shadow, boban14TVT}. 
To calculate LOS and transition probability statistics\footnote{Throughout the paper, we use the term ``LOS probabilities'' to encompass the probability of LOS, NLOSb, and NLOSv.},
we perform 
geometry-based deterministic simulations 
of LOS blockage and extract the parameters from real cities and highways for LOS blockage and transition probabilities. 
Based on the empirical results from five large cities (downtown Rome, New York, Munich, Tokyo, and London) and a highway section (10 km, including two- and three-lane per direction as well as on-ramp traffic), we perform curve fitting and obtain a set of  distance-dependent polynomial equations for both LOS and transition probabilities. To test whether the parameters extracted from a set of cities can generalize to other cities, we compare the model parameterized on the five cities to downtown Paris. The results show a high correlation for both LOS probabilities and transition probabilities. Our results can be used to generate time- and space-consistent LOS blocking for V2V channels by assigning appropriate parameters 
for each of the three states in representative urban and highway environments. %

The most important contributions of this work are as follows.
\begin{itemize}
\item We perform a systematic large-scale analysis of LOS blockage and transition probability using real locations of vehicles and other objects 
to arrive at a practical model for time evolution 
of V2V links
 in real urban and highway environments;
\item We incorporate 
the moving environment (vehicles) into the calculations of LOS and transition probability and show that its impact is significant in both urban and highway environment; we also show that, due to the low height of both Tx and Rx antennas, vehicle density has a strong impact on LOS and transition probabilities, thus requiring separate LOS blockage analysis for low, medium, and high density of vehicular traffic.
\item We extract parameters that can be used to perform efficient simulations of time-evolved V2V channels, without the need
for complex geometry-based deterministic modeling. 
\end{itemize}

The rest of the paper is organized as follows. Section~\ref{sec:RelWork} discusses existing work on spatially consistent V2V channels. Section~\ref{sec:setup} describes the model for time evolution of V2V links and the tools we use for estimating the LOS and transition probabilities. Section~\ref{sec:results} shows the LOS and transition probability parameterization results, whereas Section~\ref{sec:Discussion} discusses model validation, usage, and comparison with a state of the art model. 
Section~\ref{sec:Conclusions} concludes the paper.

\section{Related Work}\label{sec:RelWork}
While geometry-based deterministic models intrinsically support time evolution and spatial consistency, stochastic models, (either geometry-based or not) need additional mechanisms to support these features. 
While LOS probability has been extensively explored in the literature for cellular systems (for example, see~\cite{series2009guidelines}), the first steps to achieve time- and space-consistent models have been discussed only recently in the community~\cite{5GChannelModels}. For V2X and other dual-mobility communication systems such as device-to-device (D2D) communication, this is an even more challenging problem, since mobility on both sides of the link makes the modeling more complicated.

Markov chains are often used to efficiently model the time-dependent evolution of different systems.
They have been used to characterize different aspects of wireless channels. 
Gilbert-Elliot burst noise channel, a two state hidden Markov model, has been used to model the bit error probabilities for a wireless channel~\cite{wang1995finite},\cite{Ebert99agilbert}. Similarly, modeling time evolution of V2V links using Markov chains has been previously explored in literature. Dhoutaut et al.~\cite{dhoutaut06} propose a shadowing-pattern model for V2V links that uses two-state Markov chains to determine the level of shadowing caused by other vehicles on an urban street~\cite{dhoutaut06}. Similarly, Abbas et al.~\cite{abbas2012shadow} use the same model to quantify the shadowing effect of obstructing vehicles on V2V links in car following scenario in highway and urban environments. Both studies employ a small set of measurements to validate the model. 
Wang et al.~\cite{wang2014simulating} extend the Markov model to include V2V link cross-correlation (correlation of two geographically close V2V links). 

However, while the above papers employ Markov chains to model time evolution of V2V, they are based on limited measurement or simulation data applicable to a single scenario. To the best of our knowledge, our work is the first to perform a large-scale study to realistically parameterize probability of being in each of the three states (LOS, NLOSv, NLOSb) as well as the transition probabilities in both urban and highway environments. 

The whitepaper by key 3GPP stakeholders~\cite{5GChannelModels} noted that the missing components in 3GPP standardized channel models, both above and below 6~GHz, include: 
i) spatially consistent LOS probability/existence; ii) blockage modeling; and iii) moving environment (e.g., cars). This paper provides a model that contributes to the solution of these shortcomings for V2V communication.

\section{Model for Time Evolution of V2V Channels}\label{sec:setup}
There exist measurement studies that analyze the LOS blockage of V2V links using video recordings collected during the measurements (e.g.,~\cite{meireles10},\cite{boban14_TMC},\cite{abbas2015measurement}). However, extracting a general model for LOS blockage evolution using the limited amount of data collected therein is not possible, since the data does not encompass the blockage behavior across an entire environment (be it city or highway) or for different traffic conditions.
Additionally, such studies may be imprecise, because they rely on estimation of LOS blockage inferred from videos. 

In this work, we resort to a large-scale simulation study, which enables us to analyze V2V links between large number of vehicles in various environments and with different vehicle densities. 
To model the time evolution of V2V links, we apply a three-state discrete-time Markov chain, as shown in Fig.~\ref{fig:MarkovChain}. This is in contrast to state of art where two states (LOS and NLOS) are usually assumed for V2V channels. 
The probability of the states and the transition probabilities are parameterized through extensive simulation.

\subsection{Mobility modeling}
In order to obtain realistic V2V LOS blockage behavior from simulations, vehicles need to move in a realistic way over real roads. For this reason, we used SUMO (Simulation of Urban MObility)~\cite{sumo} to generate vehicular mobility in cities and on highway. By using real roadways and traffic rules and employing vehicular traffic dynamics models such as car-following~\cite{rothery92}, SUMO is capable of generating accurate vehicle positions, speeds, inter-vehicle distance, acceleration, overtaking and lane-changing maneuvers, etc. 
Previous work (e.g.,~\cite{dhoutaut06,boban14TVT}) has shown that V2V channel characteristics are affected by the density of vehicular traffic. To that end, 
for each environment, we generated three traffic densities, qualitatively characterized as low, medium and high. 
The densities in urban areas were generated according to values proposed by Ferreira et al.~\cite{ferreira09}, whereas for highways we use traffic flow measurement values reported by Wisitpongphan et al.~\cite{wisit07}. 
To allow the vehicular traffic to move into steady state, we run the mobility model for 300 seconds in each environment before performing the LOS blockage analysis. We used one second as time step for mobility simulation.

\subsection{Datasets used for training the model} 
\subsubsection{Urban}
To determine LOS and transition probabilities in urban environments, we extracted the roadway (used for mobility simulation) and object outlines (buildings, trees, walls, etc.) from OpenStreetMap~\cite{openstreetmap} for downtown areas of five cities: downtown Rome, New York, Munich, Tokyo, and London. The reason for using multiple cities is to obtain a set of ``generic'' urban parameters, i.e., those that can readily describe a typical urban environment in terms of LOS blockage, mobility patterns, road configurations, etc. The criteria for selecting the cities were practical: we selected those cities where the object outlines available in OpenStreetMap are as complete as possible. 
Additionally, we analyzed whether the vehicular mobility simulation generated by SUMO contained any anomalies (e.g., complete gridlocks due to incorrect intersection configurations or disconnected road segments). From the initially considered 10 large cities, we selected downtown Rome, New York, Munich, Tokyo, and London (along with Paris, used for testing the model) based on the described criteria. In cities, all simulated vehicles were personal cars. More details on the locations used can be found in Table~\ref{tab:Environments}.

\begin{figure}[!t]
  \begin{center}
	\includegraphics[trim=0cm 0cm 0cm 0cm,clip=true,width=0.5\textwidth]{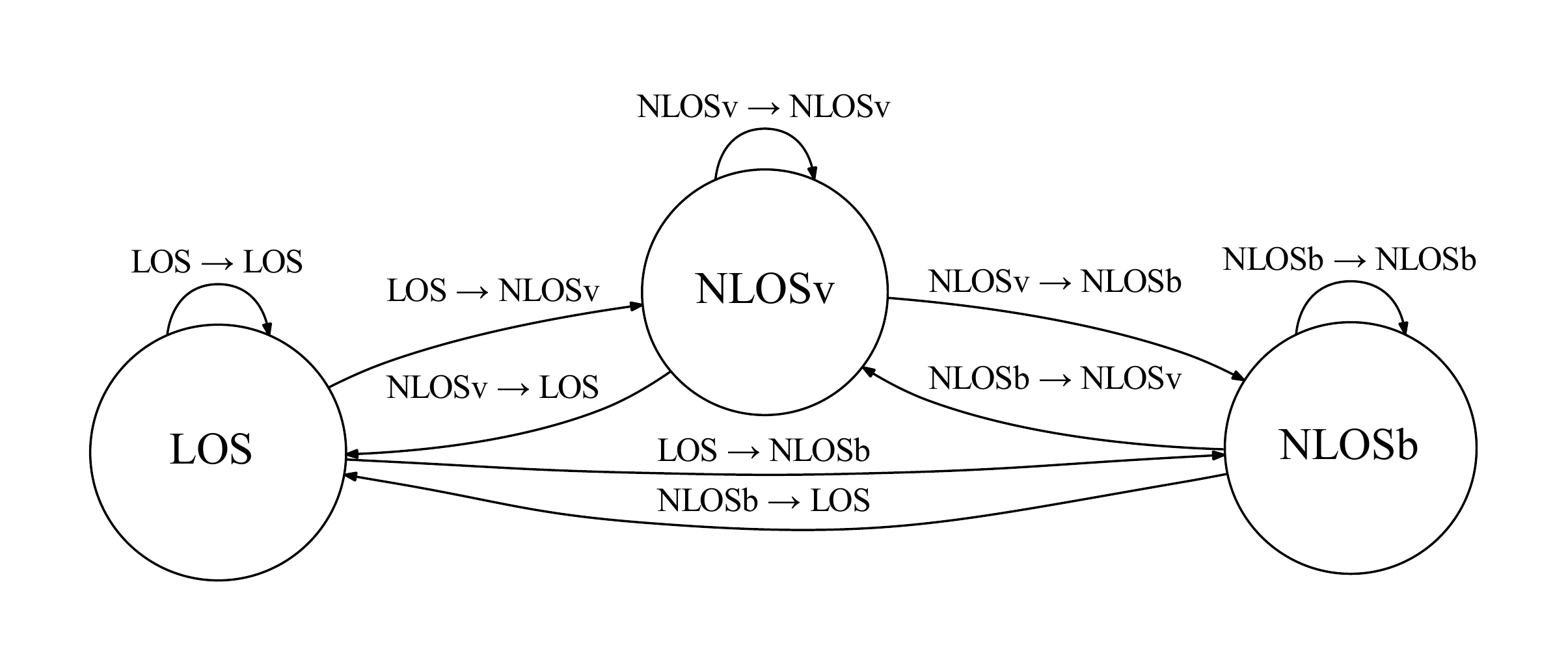}
  \end{center}
 \vspace{-1cm}
\caption{Markov chain for modeling the time evolution of V2V links. States: Line-of-Sight (LOS); ii) non-LOS due to static objects (NLOSb); and iii) non-LOS due to mobile objects (NLOSv). 
}
\label{fig:MarkovChain}
\vspace{-0.5cm}
\end{figure}

Note that, while the areas used for modeling are of limited size, for high vehicle density there were up to a hundred thousand V2V communication pairs per time step in the urban area since, for a given vehicle, all other vehicles within 500~m radius were considered as possible V2V communication pairs. For medium and high vehicular density, we perform the simulations for 300 seconds. For low density, the number of vehicles and therefore V2V links was smaller, which required 1000-second simulation runs to obtain a representative sample size.
\subsubsection{Highway}
We performed highway simulations with SUMO on a 10~km stretch of A6 highway between Braunsbach and Wolpertshausen in Bavaria, Germany. The road alternates between two and three lanes per direction and has several entry/exit ramps. We simulated the traffic so that 80\% of the vehicles enter and exit at the each end of the highway, whereas remaining 20\% enter and exit over two ramps near the beginning and the end of the stretch. Since vehicle speeds and dimensions affect the LOS blockage results, 10\% of the simulated vehicles were trucks and remaining were personal cars.
Some buildings and forest exist on each side of the highway; they can block LOS for larger Tx-Rx distances, particularly for the vehicles moving on the entry/exit ramps. For medium and high density, we perform the simulations for 1000 seconds. For low density, we perform simulations for 2000 seconds.
Note that different highways will have different configurations in terms of lane numbers, surrounding objects, etc, which will affect LOS blockage. 
It is difficult to encompass all different variations, therefore we focus on the most common highway configuration.

\begin{table}[t!]
\centering
\begin{footnotesize}

\caption{Locations and vehicle densities used for analysis} 
\label{tab:Environments}
\begin{tabular}{l c c c}
\multicolumn{4}{c}{\textbf{Environments}}\\ \hline \hline
\textbf{Location} & \multicolumn{2}{c}{\textbf{Boundary  (Latitude,Longitude)}} &  \textbf{Size} \\
& \textbf{Lower left} &  \textbf{Upper right} & \\ \hline 
\textbf{Rome} & 41.8896,12.4751 & 41.9011,12.5016 & 2.8 km$^2$ \\
\textbf{New York} & 40.7274,-74.0068 & 40.7397,-73.9764 & 3.5 km$^2$ \\
\textbf{Munich} & 48.1301,11.5553 & 48.1431,11.5928 & 4 km$^2$ \\
\textbf{Tokyo} & 35.6540,139.7256 & 35.6673,139.7561 & 4 km$^2$ \\
\textbf{London} & 51.5105,-0.1166 & 51.5230,-0.0794 & 3.6 km$^2$\\ \hline
\textbf{A6 Highway} & 49.1715,9.7441 & 49.1846,9.8630 & 10 km\\ \hline
\multicolumn{4}{c}{\textbf{Test data set for urban environment}}\\
\textbf{Paris} & 48.8489,2.3161 & 48.8575,2.3439 & 2 km$^2$ \\ [+5pt]

& \multicolumn{3}{c}{\textbf{Vehicle Densities}}\\ \hline \hline
\textbf{Environment} & \textbf{Low} &  \textbf{Medium} & \textbf{High} \\ \hline
\textbf{Urban~\cite{ferreira09}} & 24 veh/km$^2$ & 120 veh/km$^2$ & 333 veh/km$^2$\\
\textbf{Highway~\cite{wisit07}} & 500 veh/hr/dir. & 1500 veh/hr/dir. & 3000 veh/hr/dir.\\  
\end{tabular}
\end{footnotesize}
	\vspace{-0.4cm}
\end{table}

\subsection{LOS blockage analysis}
We start our analysis by acknowledging that V2V links can have their LOS blocked by two distinct object types, static and mobile, which have distinct impact on V2V links~\cite{boban14TVT}. Furthermore, static objects such as buildings, trees, etc., typically block the LOS for V2V links between vehicles that are on different roads (e.g., perpendicular roads joined by intersections). On the other hand, mobile objects (predominantly other vehicles) block the LOS over the surface of the road. 

We use the LOS blockage classification provided by GEMV$^2$, a freely available, geometry-based V2X propagation modeling tool~\cite{boban14TVT}. GEMV$^2$ uses the outlines of vehicles, buildings, and foliage to distinguish between LOS, NLOSv, and NLOSb links. In order to do so, GEMV$^2$ performs geometry-based deterministic LOS blockage analysis using the outlines of buildings and foliage from OpenStreetMap~\cite{openstreetmap} and vehicular mobility traces from SUMO~\cite{sumo}. Provided that realistic mobility model is used and the OpenStreetMap database contains majority of buildings and foliage, GEMV$^2$ produces highly realistic LOS blockage results.

In simulations, we place the antenna in the middle of the roof on vehicles and set its height to 10~cm. For a vehicle to block the V2V link (i.e., for NLOSv blockage to occur), it needs to be within 60\% of first Fresnel zone between the antennas on the communicating vehicles. 
 When the LOS is blocked by both static and mobile objects, we classify this as NLOSb state, because in vast majority of cases static objects such as buildings are the dominant blocking factor. 
 bin. 

When determining LOS, GEMV$^2$ takes into account the electromagnetic interpretation of LOS, wherein it calculates whether 60\% of first Fresnel ellipsoid is free of any obstructions, in order to determine whether LOS is blocked. That said, we evaluated the LOS blockage results for frequencies between 2~GHz and 6~GHz and the results do not differ significantly. The reason is that the Tx-Rx distances are up to 500~meters, and the resulting difference between frequencies is small; for example, the largest difference between 60\% of first Fresnel zone between 2~GHz and 6~GHz at 500~m Tx-Rx distance is 1~meter. The results showed in the paper refer to 2~GHz carrier frequency. 

\begin{figure*}[!t]
  \begin{center}
	\subfigure[Low Density.]{\label{fig:lowUrban}\includegraphics[trim=0cm 7cm 0cm 7cm,clip=true,width=0.25\textwidth]{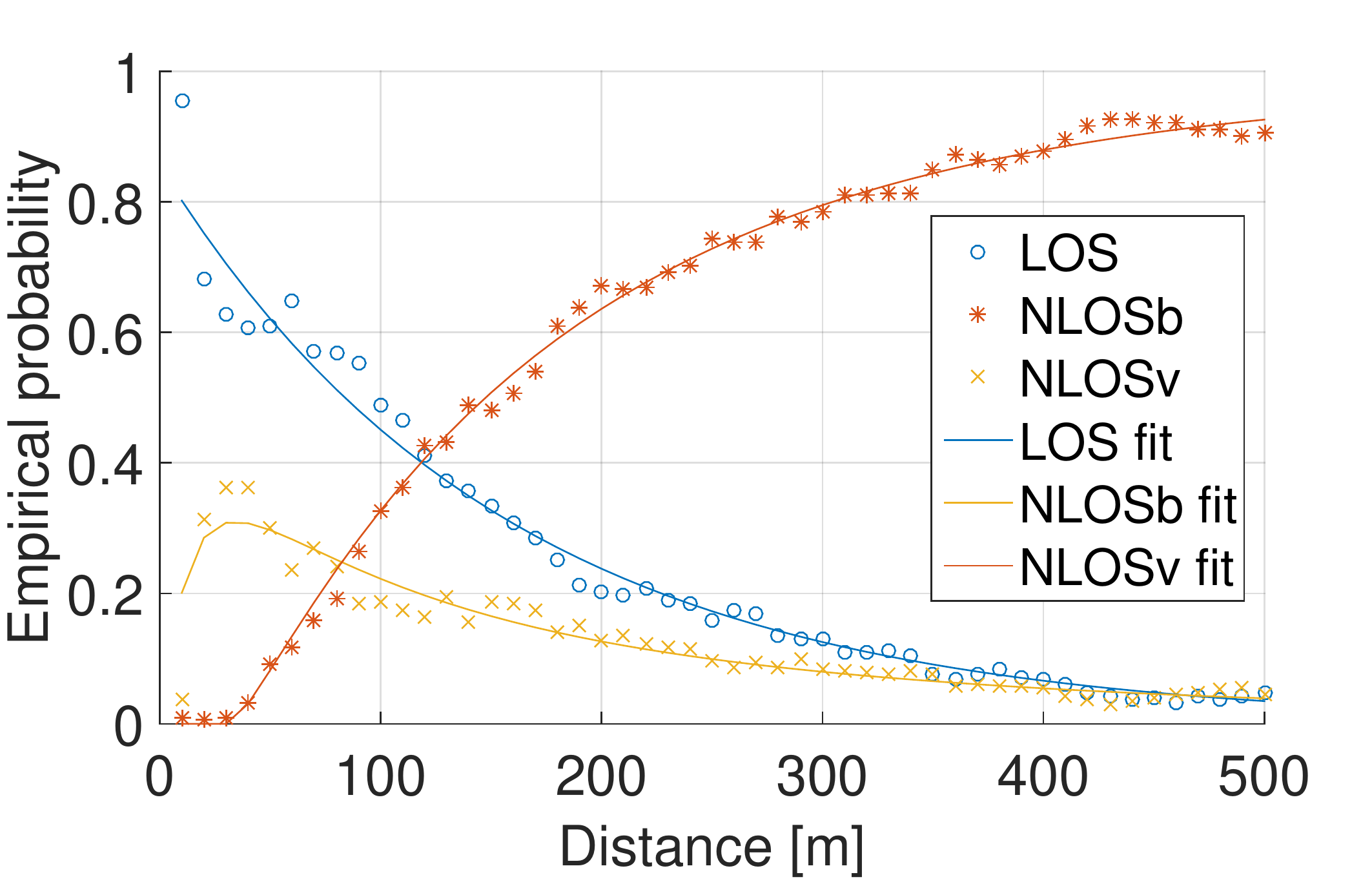}}
  	\hspace{1mm}
	\subfigure[Medium Density.]{\label{fig:medUrban}\includegraphics[trim=0cm 7cm 0cm 7cm,clip=true,width=0.25\textwidth]{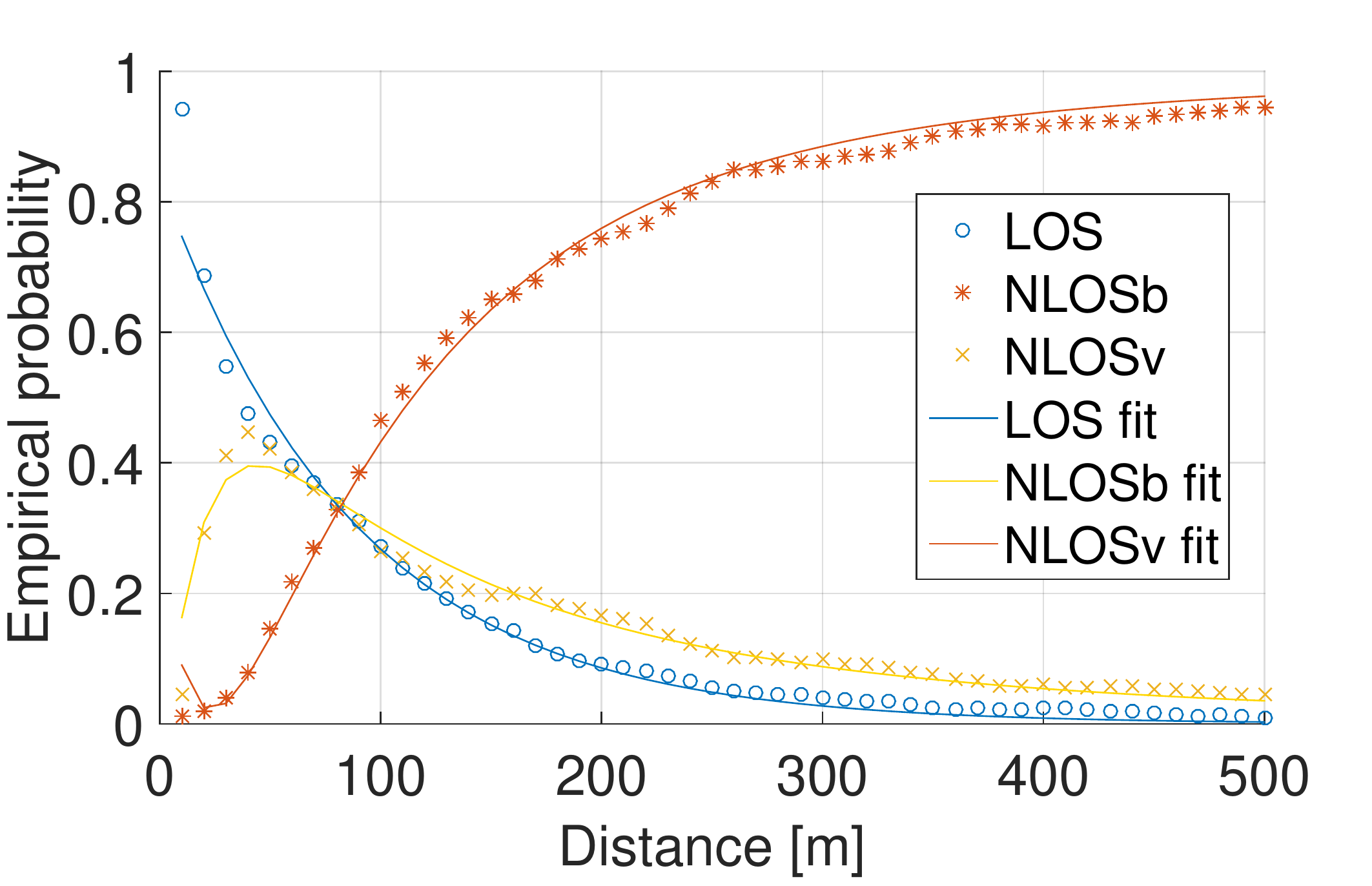}}
  	\hspace{1mm}  	
	\subfigure[High Density.]{\label{fig:highUrban}\includegraphics[trim=0cm 7cm 0cm 7cm,clip=true,width=0.25\textwidth]{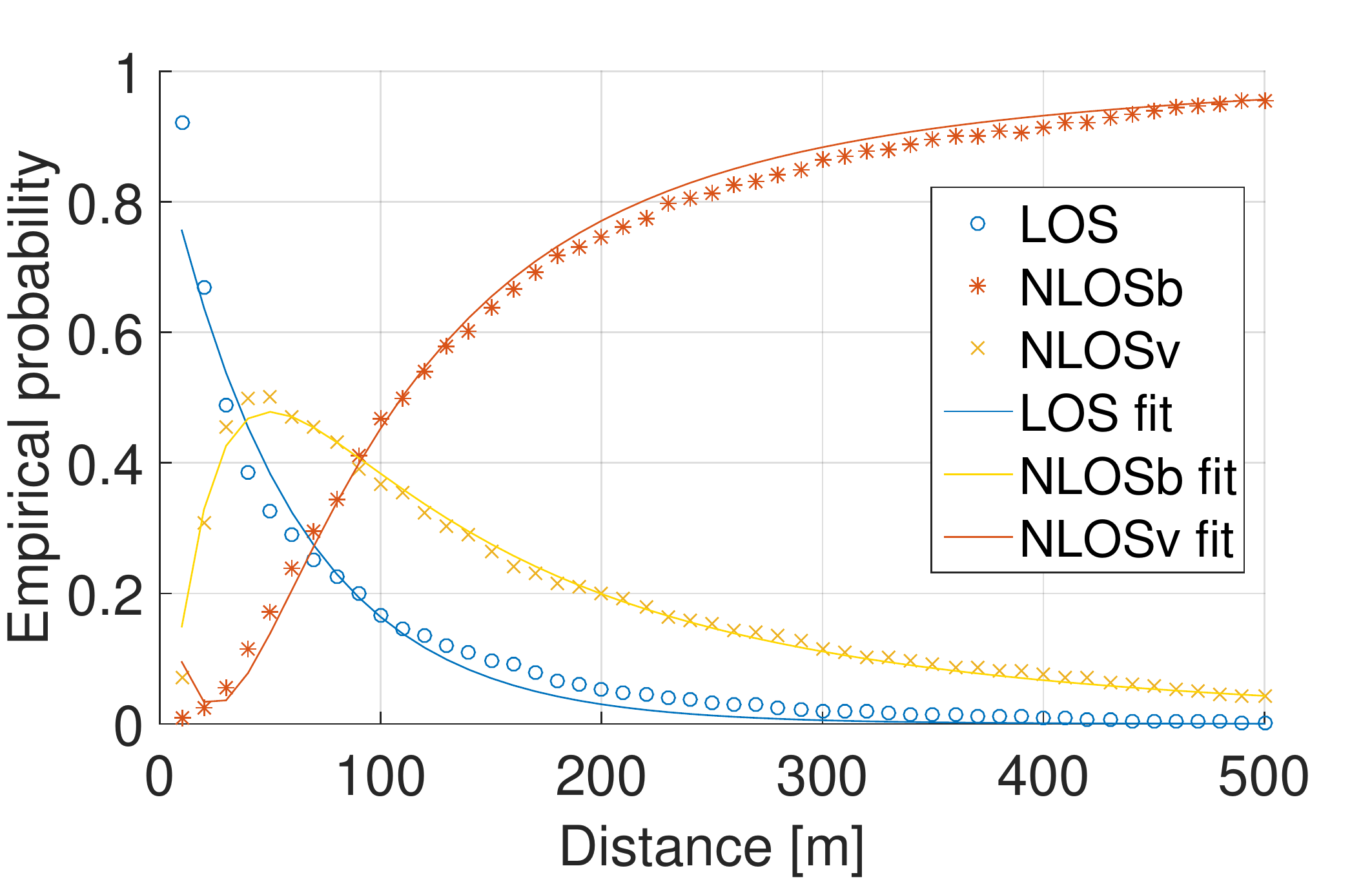}}
  \end{center}
    \vspace{-0.3cm}
\caption{LOS probabilities in urban environment: combined Rome, New York, Munich, Tokyo, and London results.}
 \vspace{-0.2cm}
\label{fig:urbanLOS}
\end{figure*}

\begin{figure*}[!t]
  \begin{center}
	\subfigure[Low Density.]{\label{fig:lowHighway}\includegraphics[trim=0cm 7cm 0cm 7cm,clip=true,width=0.25\textwidth]{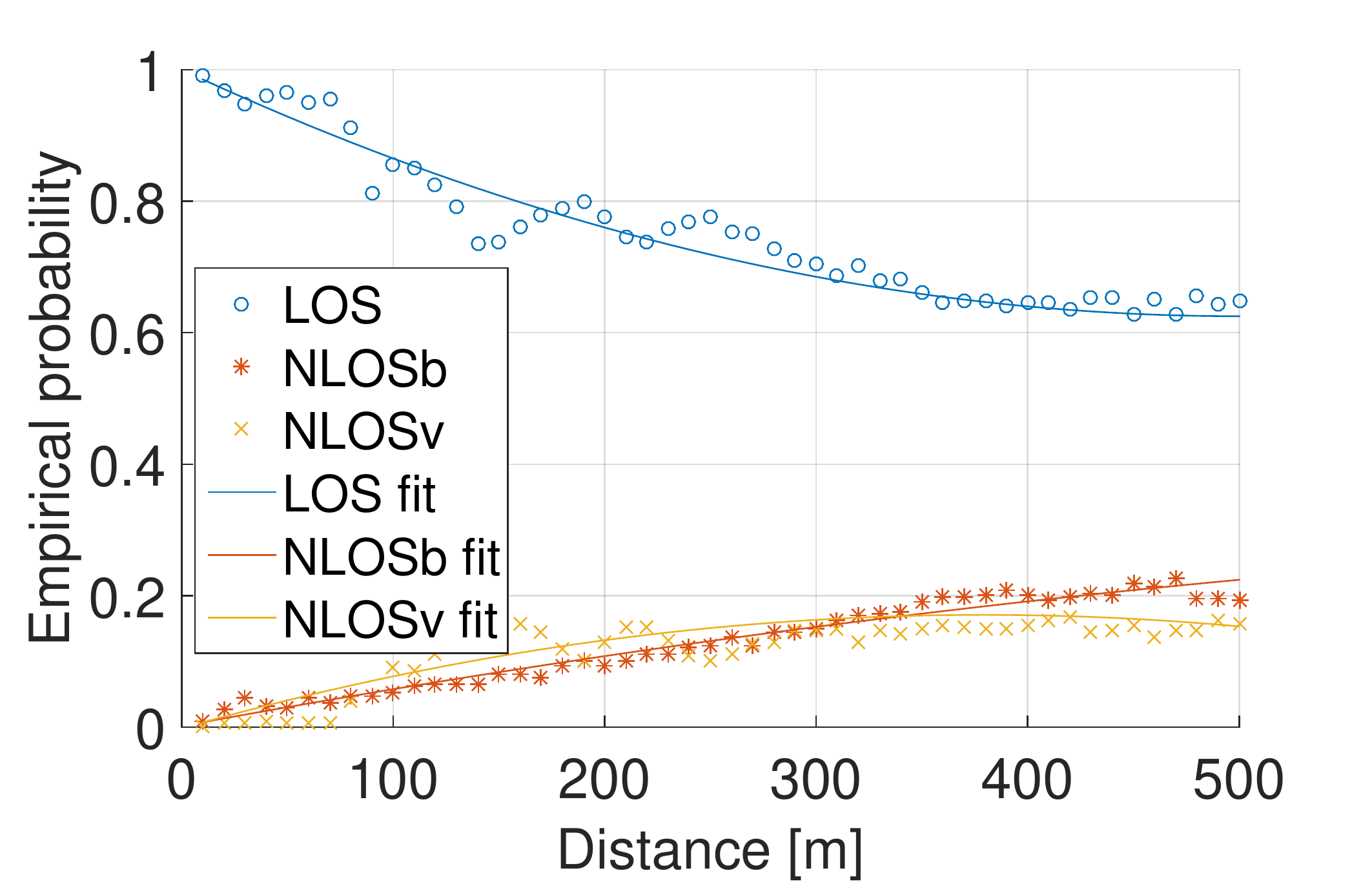}}
  	\hspace{1mm}
	\subfigure[Medium Density.]{\label{fig:medHighway}\includegraphics[trim=0cm 7cm 0cm 7cm,clip=true,width=0.25\textwidth]{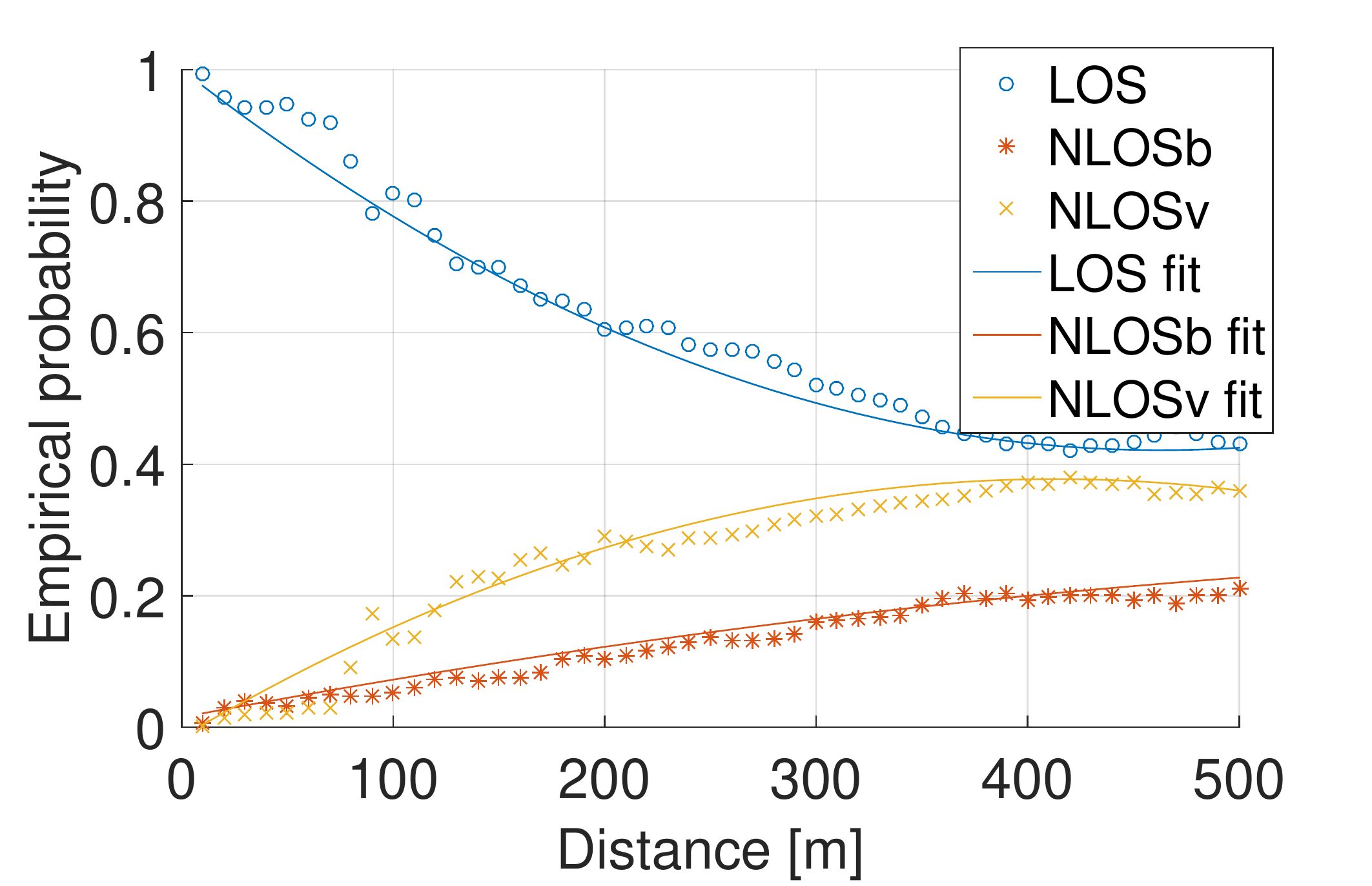}}
  	\hspace{1mm}  	
	\subfigure[High Density.]{\label{fig:highHighway}\includegraphics[trim=0cm 7cm 0cm 7cm,clip=true,width=0.25\textwidth]{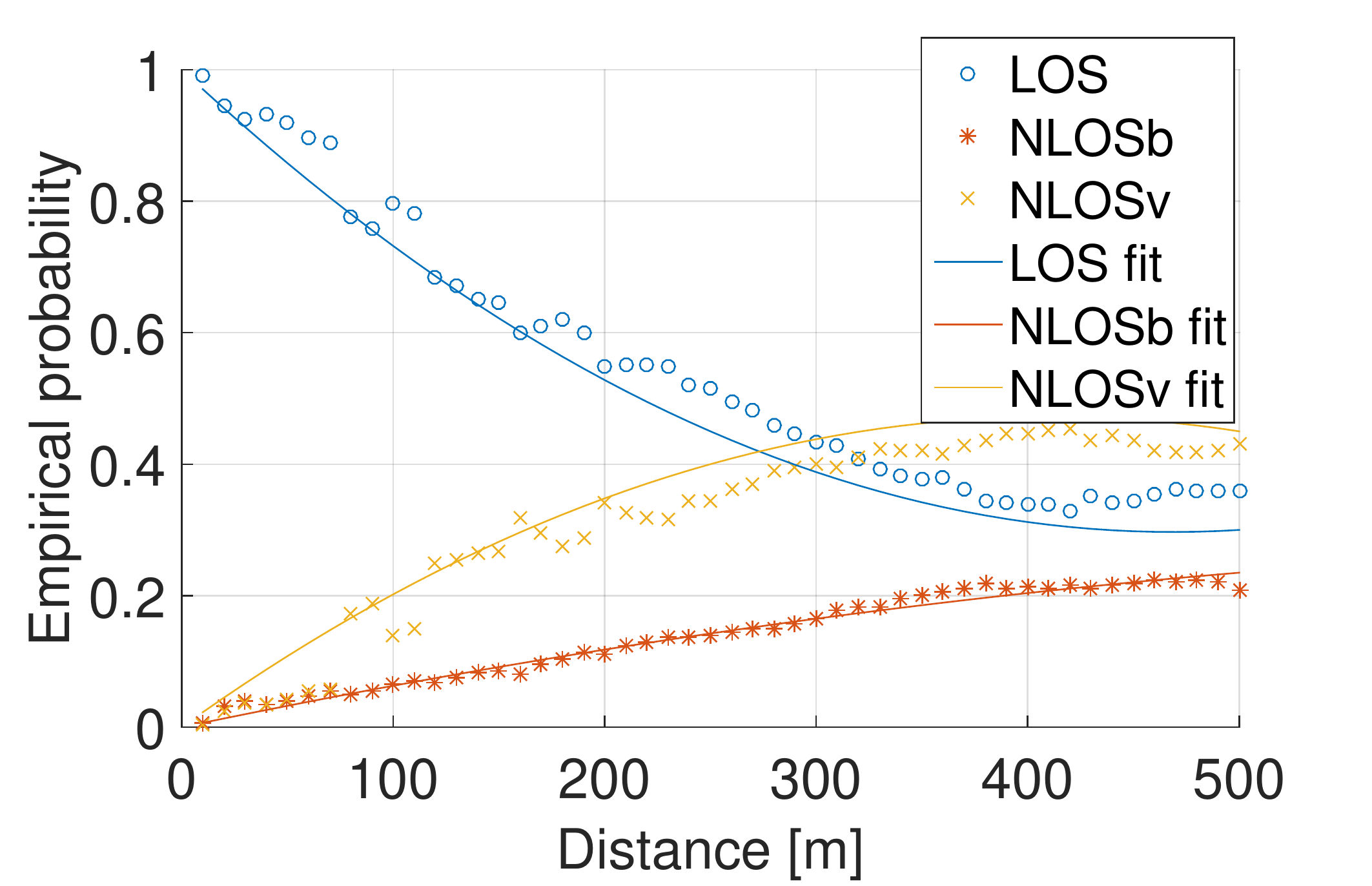}}
  \end{center}
    \vspace{-0.3cm}
\caption{LOS probabilities on A6 highway.}
 \vspace{-0.2cm}
\label{fig:highwayLOS}
\end{figure*}

\begin{figure*}[!t]
  \begin{center}
	\subfigure[Low Density.]{\label{fig:lowUrbanTrans}\includegraphics[trim=0cm 7cm 0cm 7.5cm,clip=true,width=0.25\textwidth]{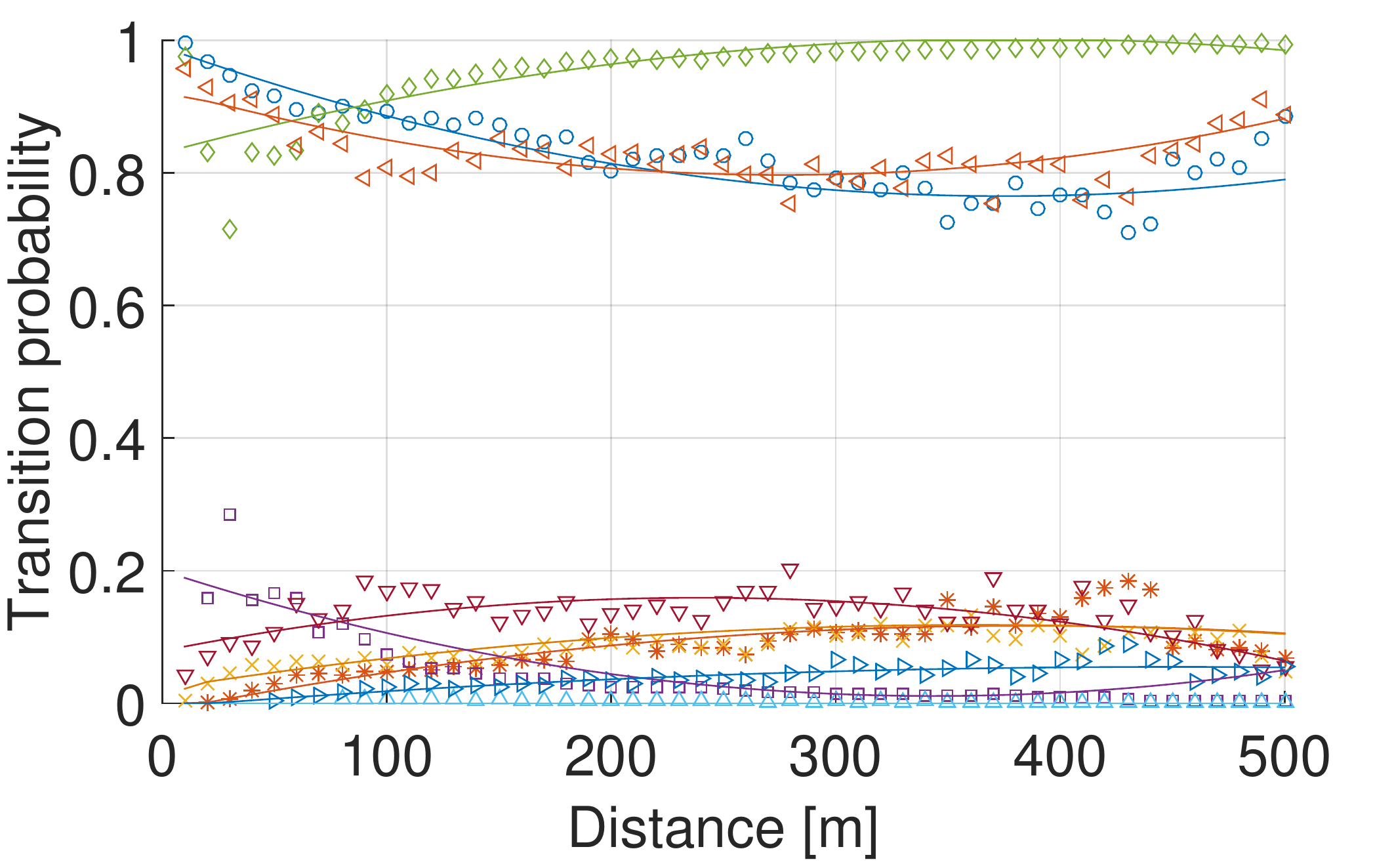}}
  	\hspace{1mm}
	\subfigure[Medium Density.]{\label{fig:medUrbanTrans}\includegraphics[trim=0cm 7cm 0cm 7.5cm,clip=true,width=0.25\textwidth]{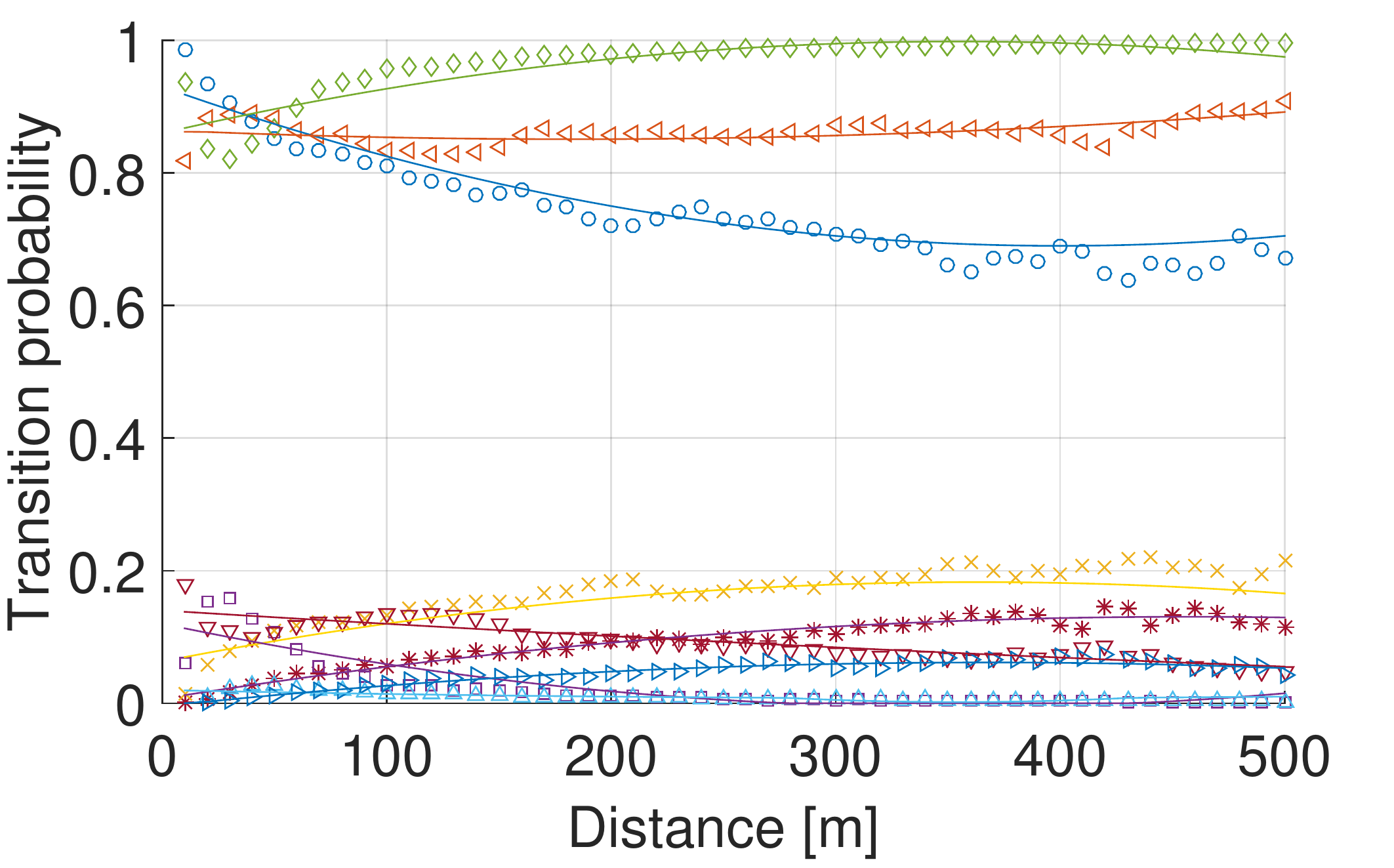}}
  	\hspace{1mm}  	
	\subfigure[High Density.]{\label{fig:highUrbanTrans}\includegraphics[trim=0cm 7cm 0cm 7.5cm,clip=true,width=0.25\textwidth]{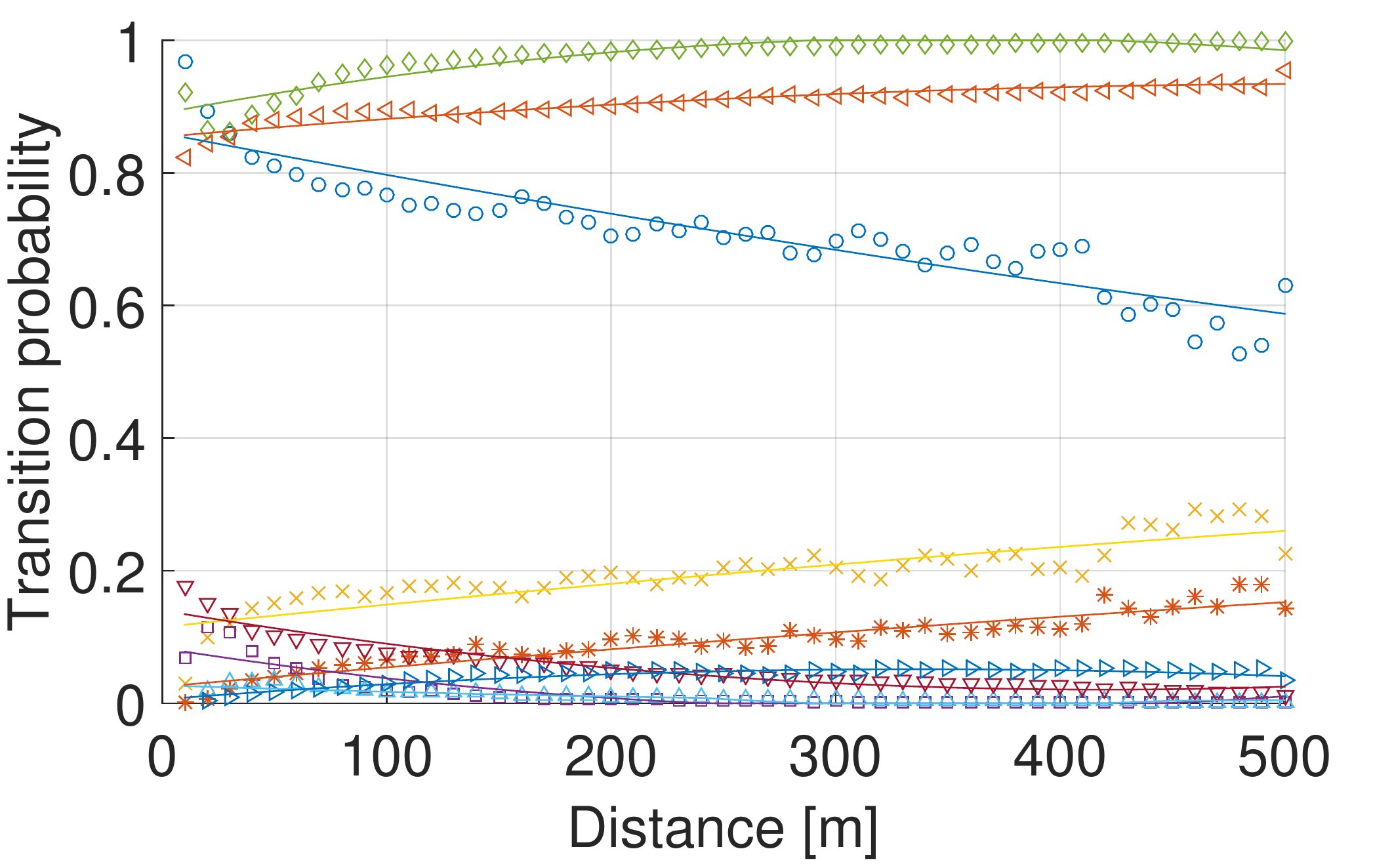}}
	\subfigure{\includegraphics[trim=3cm 27.5cm 0cm 30.5cm,clip=true,width=0.95\textwidth]{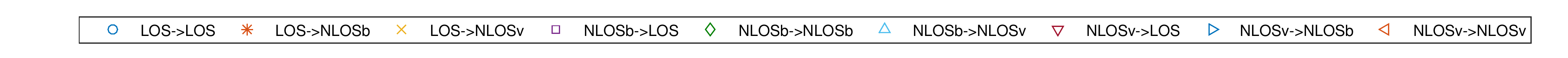}}	
  \end{center}
    \vspace{-0.3cm}
\caption{Transition probabilities in urban environment: combined Rome, New York, Munich, Tokyo, and London results.}
 \vspace{-0.2cm}
\label{fig:urbanTransProb}
\end{figure*}

\begin{figure*}[!t]
  \begin{center}
	\subfigure[Low Density.]{\label{fig:lowHwTrans}\includegraphics[trim=0cm 7cm 0cm 7cm,clip=true,width=0.25\textwidth]{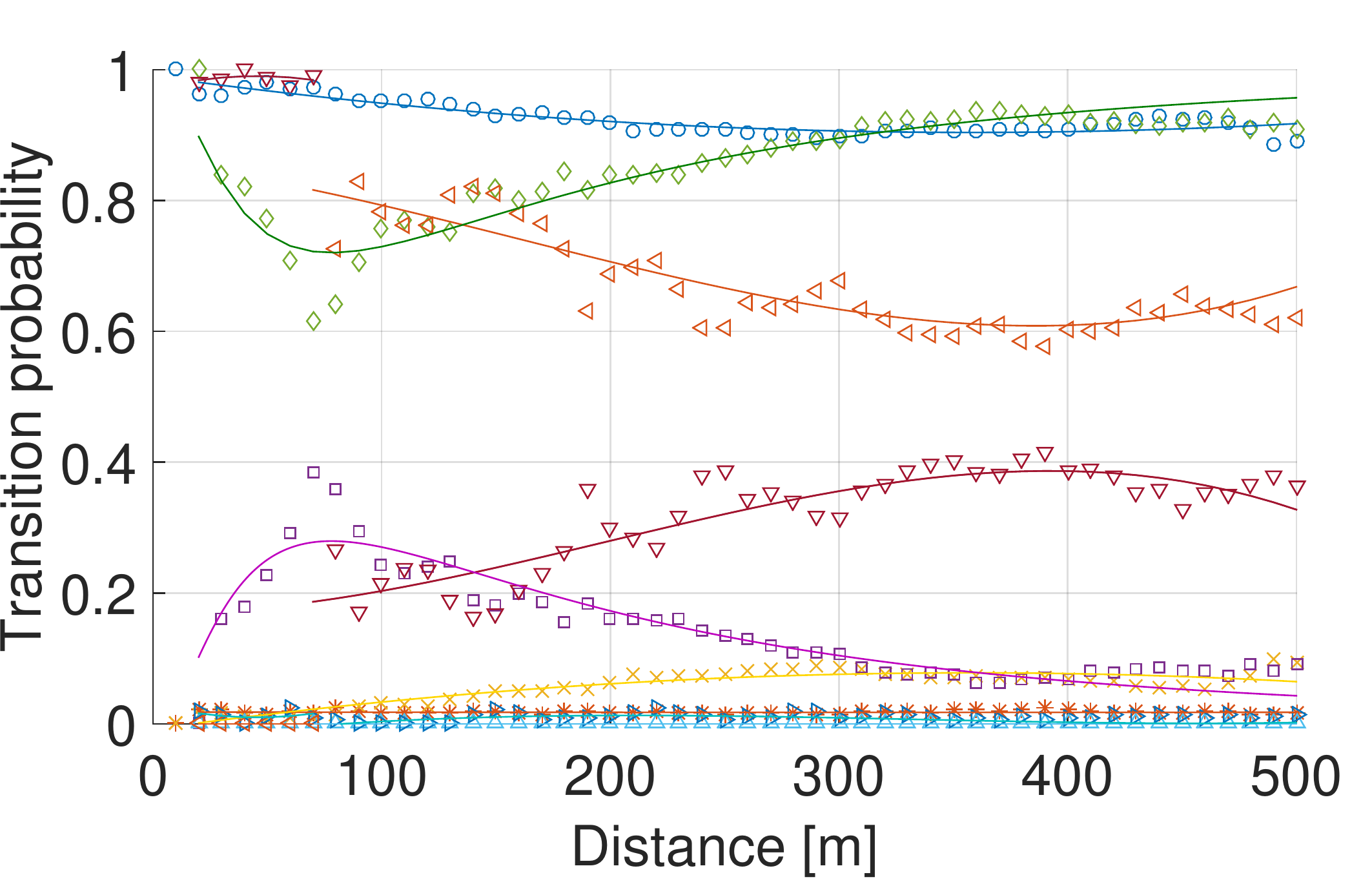}}
  	\hspace{1mm}
	\subfigure[Medium Density.]{\label{fig:medHwTrans}\includegraphics[trim=0cm 7cm 0cm 7cm,clip=true,width=0.25\textwidth]{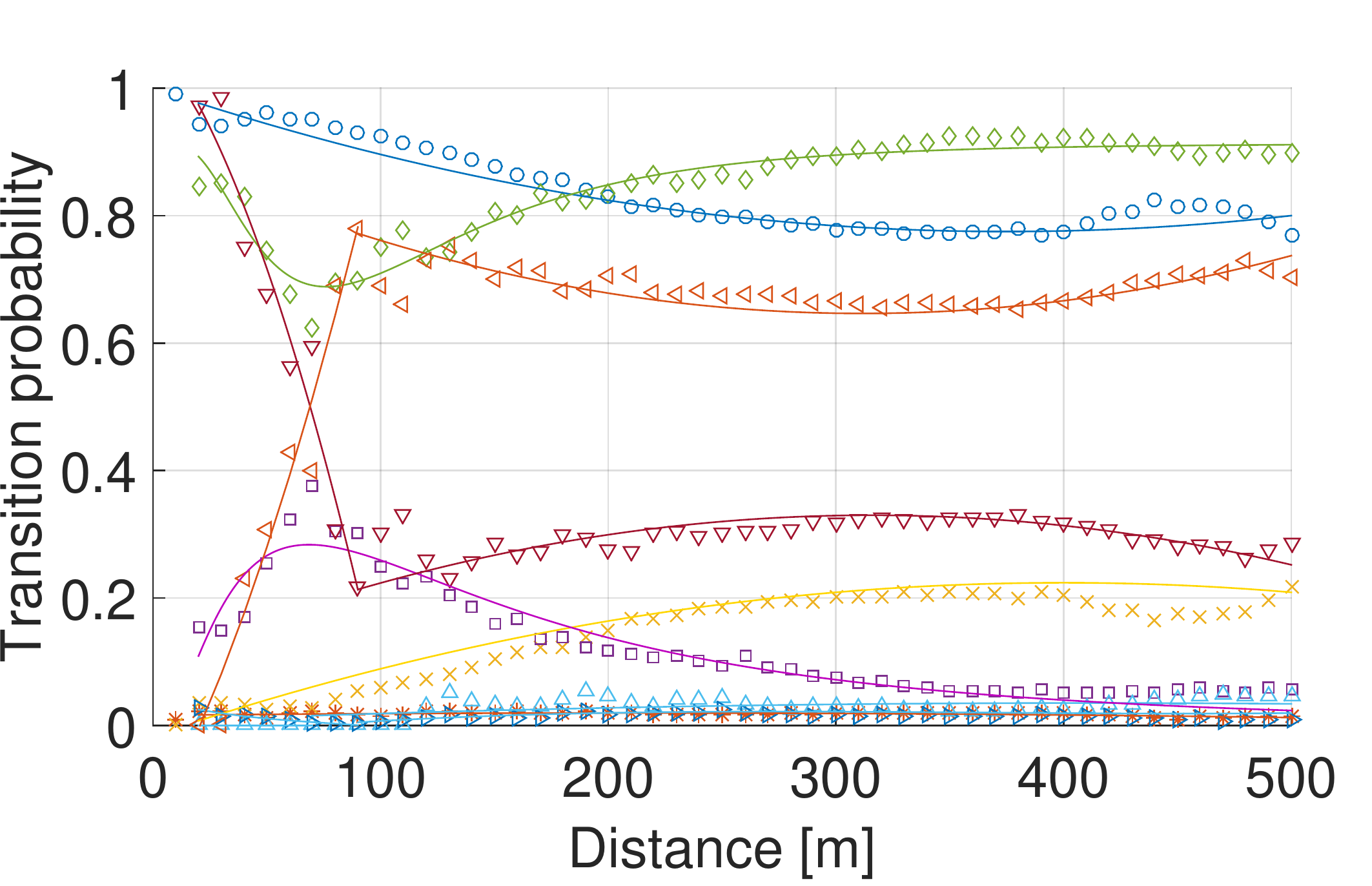}}
  	\hspace{1mm}  	
	\subfigure[High Density.]{\label{fig:highHwTrans}\includegraphics[trim=0cm 7cm 0cm 7cm,clip=true,width=0.25\textwidth]{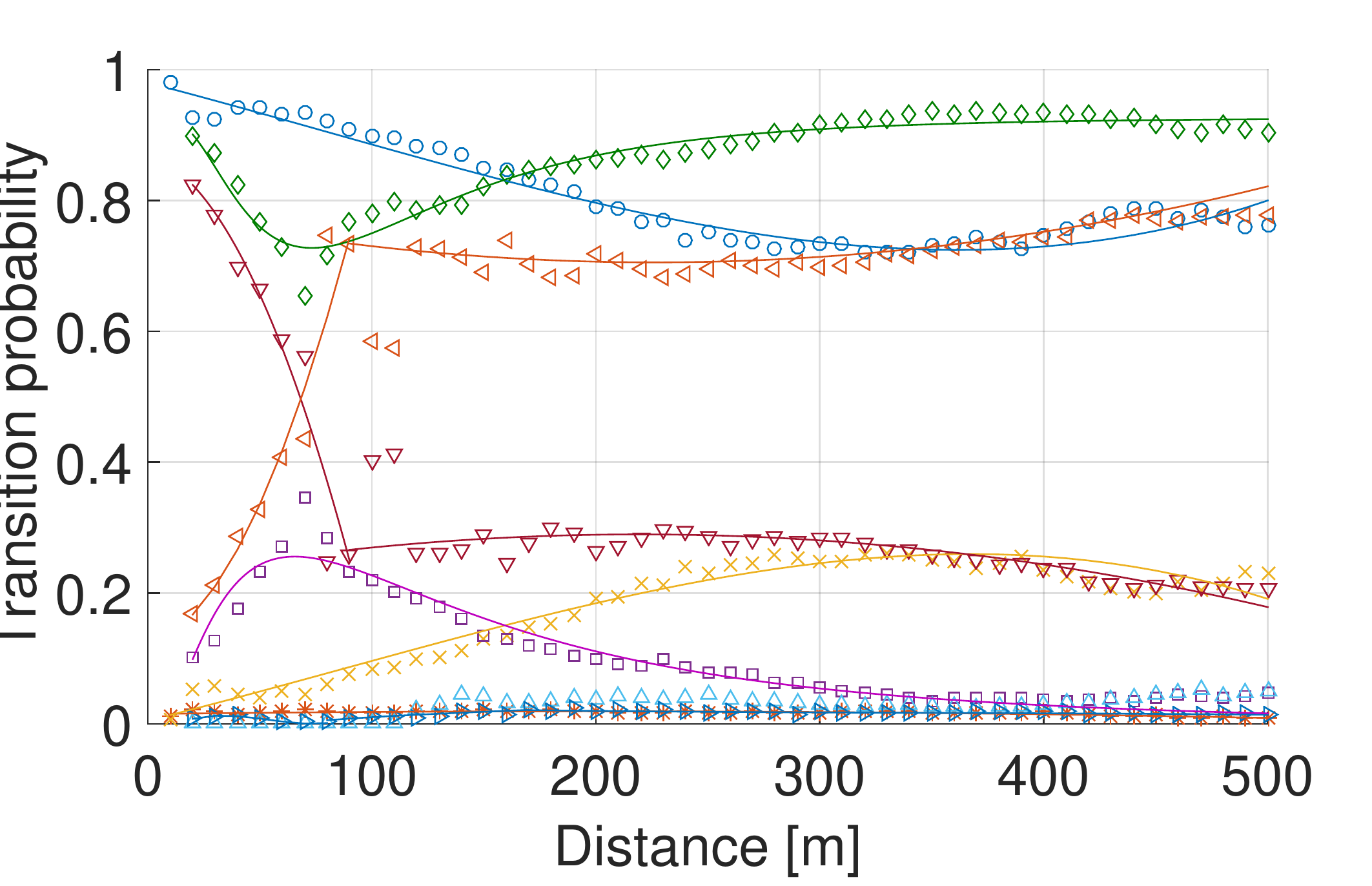}}
	\subfigure{\includegraphics[trim=3cm 27.5cm 0cm 30.5cm,clip=true,width=0.95\textwidth]{TransProbcfitlegend.pdf}}
  \end{center}
  \vspace{-0.3cm}
\caption{Transition probabilities on A6 highway.}
 \vspace{-0.2cm}
\label{fig:highwayTransProb}
\end{figure*}

\subsection{Transition probability analysis}
In order to obtain the transition probability in
the three-state Markov chain shown in Fig.~\ref{fig:MarkovChain},
we separate the data based on Tx-Rx distance into 10-meter distance bins and generate the transition probabilities for each of them. For any single step transition between states $a$ and $b$, we have the following transition probability from measurement time step $t$ to $t+1$ and for a given distance bin $d$:
\begin{align}
p_{ab}(d) = \Pr(X_{t+1}=b\mid X_{t}(d)=a). \,
\end{align}
In effect, this creates a set of distance-dependent transition probabilities
with one transition probability matrix for each Tx-Rx distance bin.
In addition to LOS blockage, GEMV$^2$ provides information about transitions between the LOS states. We use the results from GEMV$^2$ to calculate the transition probability using a frequency-based approach: for each LOS state $a$ (where $a \in {\text \{LOS, NLOSv, NLOSb\}}$) , we count the transition from $a$ to state $b$ within a distance bin $d$ (note that $b=a$ is possible, since state $a$ can transition in itself). 
Then, we divide this number by the total number of transitions from $a$ to all possible states (i.e., total number of occurrences of $a$) to obtain the empirical transition probability:

\begin{align}
\hat{p}_{ab}(d) = \frac{\sum\limits_{t} \mathbf{1}_{(X_{t+1}=b\mid X_{t}(d)=a)}}{\sum\limits_{t}\mathbf{1}_{(X_{t}(d)=a)}}.\,
\end{align}
where $\mathbf{1}_{A}$ is the indicator of event $A$.

We use the value of time interval 
equal to one second as a good trade-off between the ability to train the model (shorter time interval would require proportionally more data to obtain representative results) and precision (within one second, in vast majority of cases, there will be at most one transition between LOS states, whereas in a longer period this might not be the case). 

\section{Model parameterization}\label{sec:results}

\subsection{LOS blockage}

Fig.~\ref{fig:urbanLOS} shows the LOS probabilities for each of the three states in urban, whereas Fig.~\ref{fig:highwayLOS} shows the results for highway. Results for different vehicle densities show an intuitively expected result: the higher the vehicle density, the more probable the NLOSv state, since in low density scenarios there simply are not that many vehicles around to block the LOS. For high density urban scenario, NLOSv probability reaches 50\% when Tx and Rx are between 30 and 70 meters apart. In terms of the relationship between the probabilities of all three states, note that NLOSb remains unaffected by the increased vehicle density. The increase in NLOSv probability is at the expense of LOS probability. 
While the NLOSb in urban is most often caused by buildings lining the roads, in highway scenarios, NLOSb is caused mainly by the forest surrounding the highway, with occasional building creating NLOSb for on- and off-ramp traffic. 

\subsection{LOS transition probabilities}
Figs.~\ref{fig:urbanTransProb} and~\ref{fig:highwayTransProb} shows the state transition probabilities for urban and highway environments, respectively, along with the curve fits according to equations presented in Table~\ref{tab:EqnsTrans}. 
Across densities in urban environment (Fig.~\ref{fig:urbanTransProb}), LOS transition probability to itself reduces with increasing distance and corresponding transition from LOS to NLOSv and NLOSb increases. The remaining transition probabilities are comparatively independent of distance, with the exception of the expected increase from NLOSb to itself with increasing distance. 

Transition probabilities in highway are more dynamic. The most interesting is the relationship between NLOSv-to-NLOSv and NLOSv-to-LOS transitions: with distance increasing from zero to 100 meters, the NLOSv-to-NLOSv rapidly increases, with symmetric decrease in NLOSv-to-LOS transitions. This is a result of car-following behavior of vehicles, where vehicles are following each other at the equivalent of 1-second gap, resulting in 30 to 50-meter distance, thus making NLOSv-to-NLOSv transition increasingly more likely as distance increases. Above 100 meters, NLOSv-to-NLOSb transitions also come into play, limiting the NLOSv-to-NLOSv increase. Additionally, between 60 and 100 meters, there is a visible spike in NLOSv-to-NLOSv transitions, accompanied by a dip in NLOSv-to-LOS for all vehicle densities. This is again a result of the car-following model. 
In cases when there are three or more vehicles following each other in the same lane, the middle vehicle is often blocking the LOS between the vehicle in front and behind it. 
Since the car-following is bound to continue for some time, it results is increased NLOSv probability between front and rear vehicle, which are between 60 and 100 meters apart. Note that the resulting transition probability for low density is somewhat variable because of the smaller number of data points available. 

\subsection{Curve fitting for LOS blockage and transition probabilities}

To provide a tractable model for generation of time-evolved V2V links, we perform curve fitting for LOS and transition probabilities. 
Table~\ref{tab:EqnsLOS} shows the equations of resulting LOS probability curves, whereas Table~\ref{tab:EqnsTrans} shows the equations for transition probabilities. 

For LOS probabilities in highway, the equations are in the form of second degree polynomials: $P(LOS) = \boldsymbol{a}d^2+ \boldsymbol{b}d+ \boldsymbol{c}$, where $d$ is the Tx-Rx distance. 
In urban environment, exponential and log-normal distributions are a better fit for LOS probabilities. 
Similarly, curve fits for transition probabilities predominantly take the form of second degree polynomials, with the exception of transitions involving NLOSb state in highway environment, which are better approximated by a log-normal distribution. For more complex transition probability curves in highway, 
 we perform piece-wise curve fitting, since there are notable discontinuities, particularly in case of transitions involving NLOSv (e.g., NLOSv-to-NLOSv transition: Fig.~\ref{fig:highwayTransProb}), due to the effect of in-lane car-following on LOS blocking by vehicles.

We list LOS probability equations for two out of three curves; the third one can be obtained by subtracting from one the remaining two terms. The same applies for transition probabilities: we show two outgoing probabilities from each state. Note that, due to the imperfections of the curve fitting process, in certain scenarios at low and high distances, the probability equations will result in a probability above one. For this reason, we introduce a ceiling of one for each probability. Similarly, at very low distances (e.g., below 10~meters), the summation of the LOS probability equations (and similarly, summation of outgoing transition probability equations) can amount to more than one. In these situations, we advise to use equation for one of the three probabilities, forcing the lowest of the three probabilities to zero, and subtracting the first probability from one to obtain the third probability.

\begin{table*}[t!]
\centering
\begin{footnotesize}
\caption{Curve fitting results: LOS probabilities} 
\label{tab:EqnsLOS}
\begin{tabular}{l c c c c c c c c c}
\multicolumn{10}{c}{\textbf{Highway: LOS probability $y$ vs. distance $d$: $y = \min(1, \max(0,\boldsymbol{a}d^2+ \boldsymbol{b}d+ \boldsymbol{c}))$}}\\ \hline \hline
\textbf{Density} & \multicolumn{3}{c}{\textbf{Low}} & \multicolumn{3}{c}{\textbf{Medium}} &
\multicolumn{3}{c}{\textbf{High}}\\
& \textbf{a} & \textbf{b} & \textbf{c}
& \textbf{a} & \textbf{b} & \textbf{c} 
& \textbf{a} & \textbf{b} & \textbf{c}  \\ \hline 
\\[-5pt]
\textbf{LOS} & $1.5e^{-6}$ & -0.0015 & 1 & $2.7e^{-6}$ & -0.0025 & 1  &  $3.2e^{-6}$ & -0.003 & 1 \\
\textbf{NLOSb} &   $-2.9e^{-7}$ & 0.00059 & 0.0017 & $-3.7e^{-7}  $ & 0.00061 & 0.015  &  $ -4.1e^{-7} $ & 0.00067 & 0 
\\\\

 \multicolumn{10}{c}{\textbf{Urban: LOS probability $y$ vs. distance $d, d \ge 0: y = \min(1, \max(0, f(d)))$}}\\ \hline \hline
\textbf{Density} & \multicolumn{3}{c}{\textbf{Low}} & \multicolumn{3}{c}{\textbf{Medium}} & \multicolumn{3}{c}{\textbf{High}} \\ \hline
\\[-5pt]
\textbf{LOS} & \multicolumn{3}{c}{$0.8548e^{-0.0064d}$} & \multicolumn{3}{c}{$ 
0.8372e^{-0.0114d}$} &
 \multicolumn{3}{c}{$ 0.8962e^{-0.017d}$}\\
\textbf{NLOSv}  & \multicolumn{3}{c}{$\frac{1}{0.0396d}e^{\frac{(-(\ln(d)-5.2718)^2}{3.4827}}
$}&  \multicolumn{3}{c}{$\frac{1}{0.0312d}e^{\frac{(-(\ln(d)-5.0063)^2}{2.4544}}$} & \multicolumn{3}{c}{$\frac{1}{ 0.0242d}e^{\frac{(-(\ln(d)-5.0115)^2}{2.2092}}$} \\  
\end{tabular}

\end{footnotesize}
\end{table*}

\begin{table*}[t!]
\centering
\begin{footnotesize}
\caption{Curve fitting results: Transition probabilities. } 
\label{tab:EqnsTrans}
\begin{tabular}{l c c c   c c c   c c c  }
\multicolumn{10}{c}{\textbf{Highway: Transition probability $y$ vs. distance $d$: $y = \min(1, \max(0,f(d)))$ d$_T$=70 (Low) / 90 (Medium/High)}}\\ \hline \hline
\textbf{Density} & \multicolumn{3}{c}{\textbf{Low}} & \multicolumn{3}{c}{\textbf{Medium}} &
\multicolumn{3}{c}{\textbf{High}}\\\\[-5pt]
\textbf{LOS $\rightarrow$ LOS} 
&  \multicolumn{3}{c}{$6.7e^{-7}d^2-4.8e^{-4}d + 0.99$}
& \multicolumn{3}{c}{$1.6e^{-6}d^2-1.2e^{-3}d+ 1$} 
&  \multicolumn{3}{c}{$2.1e^{-6}d^2-1.5e^{-3}d+ 1$} \\
\textbf{LOS $\rightarrow$ NLOSb} 
&  \multicolumn{3}{c}{$4e^{-9}d^2-2.7e^{-6}d + 0.018$}
&  \multicolumn{3}{c}{$-8.4e^{-8}d^2+3.5e^{-5}d + 0.016$}
&  \multicolumn{3}{c}{$-1.1e^{-7}d^2+4.3e^{-5}d + 0.015$}  \\ 
\hline \\[-5pt]
\textbf{NLOSb $\rightarrow$ LOS } 
&  \multicolumn{3}{c}{$\frac{1}{0.0289d}e^{\frac{(-(\ln(d)-5.2782)^2}{1.8424}}$} 
&  \multicolumn{3}{c}{$\frac{1}{0.0346d}e^{\frac{(-(\ln(d)-5.021)^2}{1.5875}}$} 
&  \multicolumn{3}{c}{$\frac{1}{0.0411d}e^{\frac{(-(\ln(d)-4.927)^2}{1.4876}}$} \\  
\textbf{NLOSb $\rightarrow$ NLOSb }
&  \multicolumn{3}{c}{$1-\frac{1}{0.0289d}e^{\frac{(-(\ln(d)-5.2782)^2}{1.8424}}$} 
&  \multicolumn{3}{c}{$0.9132-\frac{1}{0.0484d}e^{\frac{(-(\ln(d)-4.7076)^2}{0.7480}}$} 
&  \multicolumn{3}{c}{$0.9264-\frac{1}{0.056d}e^{\frac{(-(\ln(d)-4.7012)^2}{0.8186}}$} \\[+2pt]
\hline \\[-5pt]
\textbf{NLOSv $\rightarrow$ LOS }& \multicolumn{9}{c}{}\\
\hspace{0.5cm} $d \in (0, d_T)$ & 
  \multicolumn{3}{c}{$-9.8e^{-6}d^2+ 8.9e^{-4} d+ 0.97$} & 
    \multicolumn{3}{c}{$-4.8e^{-5}d^{2} - 5.62e^{-3}d + 1.11$} 
    & 
 \multicolumn{3}{c}{$-6.51e^{-5}d^{2} - 1.04e^{-3}d + 0.8706$}  
    \\   
\hspace{0.5cm}$d \in (d_T, 500):$ &  \multicolumn{3}{c}{$-2e^{-6}d^2$+ 1.6$e^{-3} d+ 0.051$} &
  \multicolumn{3}{c}{$-2.286e^{-6}d^{2} + 1.443e^{-3}d + 0.1022$}
  &
  \multicolumn{3}{c}{$-1.412e^{-6}d^{2} + 6.196e^{-4}d + 0.2216$}
\\
\textbf{NLOSv $\rightarrow$ NLOSb }& \multicolumn{9}{c}{}\\
\hspace{0.5cm} $d \in (0, d_T)$ & 
 \multicolumn{3}{c}{$9.8e^{-6}d^2- 8.9e^{-4} d+ 0.03$} &
   \multicolumn{3}{c}{$4.4e^{-6}d^{2} - 8.335e^{-4}d + 0.042$}
   &
   \multicolumn{3}{c}{$1.254e^{-7}d^{2} - 3.775e^{-5}d + 9.853e^{-3}$}
\\   
\hspace{0.5cm}$d \in (d_T, 500):$ &  
  \multicolumn{3}{c}{$-1.4e^{-7}d^2$+9.1$e^{-5} d- 0.0016$} &
     \multicolumn{3}{c}{$-2.7e^{-7}d^{2} + 1.5e^{-4}d - 0.0031$}
     &
     \multicolumn{3}{c}{$-1.4e^{-7}d^{2} + 8.3e^{-5}d - 0.0065$}
\\\\[-3pt]

 \multicolumn{10}{c}{\textbf{Urban: Transition probability $y$ vs. distance $d$: {$y = \min(1, \max(0,\boldsymbol{a}d^2+ \boldsymbol{b}d+ \boldsymbol{c}))$}}}\\ \hline \hline
\textbf{Density} & \multicolumn{3}{c}{\textbf{Low}} & \multicolumn{3}{c}{\textbf{Medium}} & \multicolumn{3}{c}{\textbf{High}} \\
& \textbf{a} & \textbf{b} & \textbf{c}
& \textbf{a} & \textbf{b} & \textbf{c} 
& \textbf{a} & \textbf{b} & \textbf{c}  \\ \hline
\textbf{LOS $\rightarrow$ LOS} &  
{$1.6e^{-6}$} & 
{$-1.2e^{-3}$} & {$0.99$} & 
 $1.5e^{-6}$ & 
{$-1.2e^{-3}$} & {$0.93$} & 
 {$2.1e^{-7}$} & 
{$-6.5e^{-4}$} & {$0.86$} \\
\textbf{LOS $\rightarrow$ NLOSb} &
 $-8.7e^{-7}$ & $6.7e^{-4}$ & -0.012 &  
 $-5.9e^{-7}$ & 
{$5.4e^{-4}$} & {$0.0069$} & 
$-9e^{-8}$ & $3e^{-4}$ & 0.025 \\
\hline \\[-5pt]
\textbf{NLOSb $\rightarrow$ LOS} & 
 {$1.6e^{-6}$} & 
{$-1.1e^{-3}$} & 
{$0.2$} &   
$1e^{-6}$ & 
{$-7.1e^{-4}$} & {$0.12$} &
 {$7.7e^{-7}$} & 
{$-5.3e^{-4}$} & {$0.083$} \\ 
\textbf{NLOSb $\rightarrow$ NLOSb} &
  $-1.2e^{-6}$ & $9.1e^{-4}$ & 0.83 & 
   $-1.1e^{-6}$ & $7.8e^{-4}$ & 0.86 
&   $-9e^{-7}$ & $6.4e^{-4}$ & 0.89\\ 
\hline \\[-5pt]
\textbf{NLOSv $\rightarrow$ LOS} & 
 {$-1.4e^{-6}$} & 
{$6.7e^{-4}$} & {$0.079$} &  
$8.1e^{-8}$ & 
{$-2.1e^{-4}$} & {$0.14$} &   
 {$6.8e^{-7}$} & 
{$-5.7e^{-4}$} & {$0.14$} \\
\textbf{NLOSv $\rightarrow$ NLOSb} &
  $-3e^{-7}$ & $2.7e^{-4}$ & -0.0059 & 
    $-4.9e^{-7}$ & $3.6e^{-4}$ & -0.0046
& $-4e^{-7}$ & $2.7e^{-4}$ & 0.0058 \\ 
\end{tabular}
\end{footnotesize}
\end{table*}

\section{Validation, Use, and Comparison to State of Art}\label{sec:Discussion}
\subsection{Validation of the proposed model}

To test how well the proposed model generalizes to other cities (i.e., those on which it was not trained), we compared the model trained on the data from five cities shown in Table~\ref{tab:Environments} and a data set obtained in downtown Paris (also described in Table~\ref{tab:Environments}). 
Table~\ref{tab:corrCoeff} shows the results of comparison in terms of Pearson Correlation Coefficient for both the LOS probabilities and transition probabilities. The correlation is above 0.95 for all three LOS probabilities. Similarly, the correlation coefficients for transition probabilities are high (average correlation equals 0.84), with only the NLOSv$\rightarrow$NLOSv below 0.8. The most likely reason for NLOSv$\rightarrow$NLOSv discrepancy is that the streets in Paris are significantly narrower and with fewer lanes per direction than in the five cities that the model was trained on. The narrower street configuration results in higher probability of continued LOS blockage by vehicles.
Overall, however, the correlation results show that the model fits well to urban environments outside those that it was trained on and can thus be used with confidence as a representative model for ``typical'' urban environments. 

\subsection{Usage of the model}

Algorithm~\ref{alg:TransProb} describes how the developed model can be used to generate LOS states for a V2V link. 
For $n$ time steps, the algorithm generates a set of $n$-tuple states by using transition equations (Table~\ref{tab:EqnsTrans}). 
The states represent the time evolution of the link in the given environment and for a given density. 
Initial state is selected randomly based on the probability of each state given the distance (Figs.~\ref{fig:urbanLOS},~\ref{fig:highwayLOS}). Since the model takes as input distance between Tx and Rx for each time step, there needs to exist consistency between subsequent distances between Tx and Rx in order to obtain credible results for LOS blockage and transition probabilities. Therefore, we analyze the distribution of speed in each of the six scenarios (highway/urban, low/medium/high density), which combined with the one-second interval results in the Tx-Rx distance variation.

\begin{figure}[!t]
\vspace{-0.3cm}
  \begin{center}
	\includegraphics[trim=0cm 8.5cm 0cm 8.5cm,clip=true,width=0.45\textwidth]{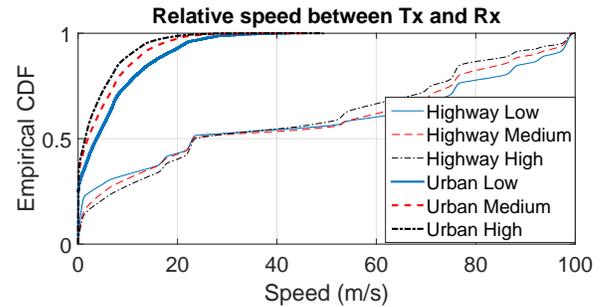}
  \end{center}
  	\vspace{-0.5cm}
\caption{Relative speed between communicating vehicles in highway and urban environments.}
\label{fig:Speed}
\vspace{-0.3cm}
\end{figure}

Fig.~\ref{fig:Speed} shows the distribution of relative speeds between Tx and Rx; the figure helps in quantifying reasonable distance that the vehicles can travel in a given environment.
In urban environments, the relative bearing of two vehicles can be between 0 and 360 degrees, whereas on highway the vehicles travel either in the same 
or opposite directions, 
with the slight deviation to this rule caused by on- and off-ramp traffic. Therefore, as shown in Fig.~\ref{fig:Speed},  
the relative speed in urban areas is mostly limited to 0-20~m/s, whereas the relative speed on highway will be distinct for same and opposite traffic, with same direction traffic ranging from 0-25~m/s 
and opposite ranging from 50-100~m/s. 
Note that the speeds can readily be translated in distance traveled, since the time step under consideration is fixed to one second. 
For considerably different vehicle speeds (e.g., in case of 
heavy traffic jams), 
the LOS blockage and transition probabilities would need to be adjusted accordingly.

\begin{table}[t!]
\vspace{-0.3cm}
\centering
\begin{footnotesize}
\caption{Correlation coefficients: Paris vs model (trained on Rome, New York, Munich, Tokyo, and London data sets). Medium density.} 
\label{tab:corrCoeff}
\begin{tabular}{l c c c}
\multicolumn{4}{c}{\textbf{LOS Probabilities}}\\ \hline \hline 
\\[-7pt]
\textbf{Paris/model} & \multicolumn{1}{c}{\textbf{LOS}} & \multicolumn{1}{c}{\textbf{NLOSb}} &
\multicolumn{1}{c}{\textbf{NLOSv}}\\
Corr. coeff. & 0.9565 & 0.9823 & 0.9528
\\\\

\multicolumn{4}{c}{\textbf{Transition Probabilities}}\\ \hline \hline
\\[-7pt]
\textbf{Paris/model} & \multicolumn{1}{c}{\textbf{LOS}} & \multicolumn{1}{c}{\textbf{NLOSb}} &
\multicolumn{1}{c}{\textbf{NLOSv}}\\
LOS $\rightarrow$ &   0.8037  &  0.9857 &   0.8319 \\
NLOSb $\rightarrow$ & 0.9587  &  0.8316 &   0.8102\\
NLOSv $\rightarrow$ & 0.8488  &  0.9080 &   0.5851
\\
\end{tabular}
\end{footnotesize}
\vspace{+0.5cm}
\end{table}

\begin{algorithm*}
\begin{footnotesize}

\caption{Algorithm for generating time- and space-consistent LOS states}\label{alg:TransProb}
\algorithmicrequire{ $env, density, LOSProb, transProb, 
numTimeSteps, distVector$} \Comment{Input: environment (urban, highway), density (low, medium, high), LOS probability equations (Table~\ref{tab:EqnsLOS}), transition probability equations (Table~\ref{tab:EqnsTrans}), number of time steps to generate, Tx-Rx distances for time steps)}\\
\algorithmicensure{ $LOSstates$}
\begin{algorithmic}[1]
\Procedure{Generate}{$LOSstates$}
\State $initialState \gets LOSProb(distVector(1),env, density)$ \Comment{Get initial state using LOS probability equations (Table~\ref{tab:EqnsLOS}) for given distance and scenario}
\State $LOSstates(1) \gets initialState$
\For{$ii=2$ to $numTimeSteps$}

\State $transMatrix \gets transProb(env,density,distVector(ii))$ \Comment{Use transition equations (Table~\ref{tab:EqnsTrans}) for selected scenario to get transition matrix}
\State $transMatrixRow \gets transMatrix(LOSstates(ii-1))$ \Comment{Select matrix row based on previous state}
\State $LOSstates(ii) \gets transMatrixRow(
randNum)$ \Comment{Generate random number weighted by transition probabilities to select next LOS state} 
\EndFor
\EndProcedure
\end{algorithmic}

\end{footnotesize}
\end{algorithm*}

\subsection{Path loss comparison with non-time-evolved model}
\begin{figure}[!t]
\vspace{-0.5cm}
  \begin{center}
	\includegraphics[trim=3cm 12.5cm 3cm 12.5cm,clip=true,width=0.38\textwidth]{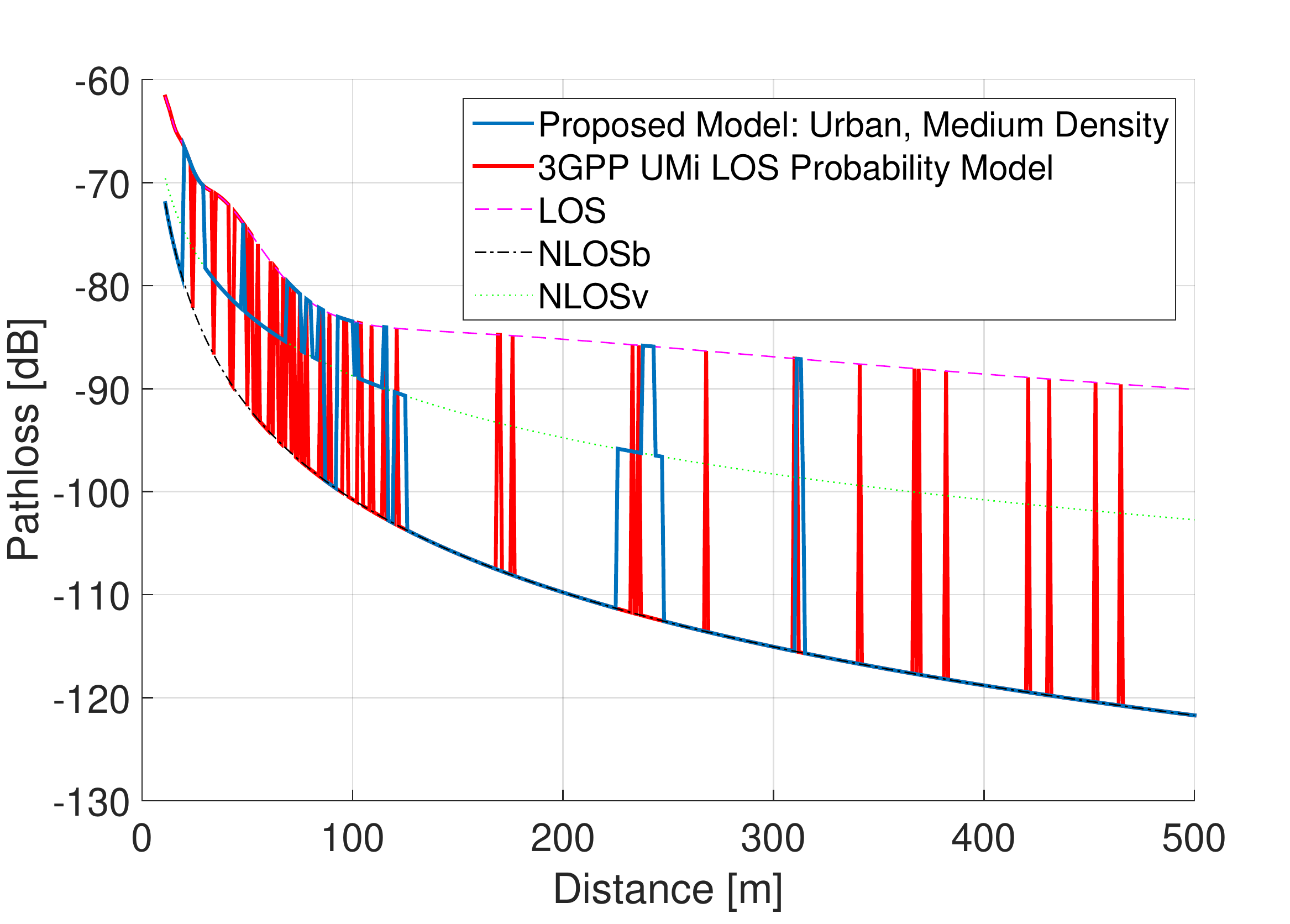}
  \end{center}
\vspace{-0.5cm}
\caption{Comparison of path loss generated by proposed model and 3GPP UMi LOS probability model for a single V2V link with Tx and Rx moving apart from 0 to 500~m with relative speed of 1~m/s. For reference, path loss for LOS, NLOSv, and NLOSb states are plotted.}
\label{fig:ComparisonPL}
\vspace{-0.3cm}
\end{figure}

Since there are no comprehensive V2V link LOS blockage and transition probability models available in the literature, we compare the proposed model with the well-established 3GPP/ITU Urban Micro (UMi) LOS probability model~\cite{series2009guidelines}, which is currently also used for D2D communication: 
\begin{align}
P(LOS) = \min{\left( \frac{d_1}{d},1\right)} \times \left(1 – e^{\frac{-d}{d_2}}\right) + e^{\frac{-d}{d_2}},
\end{align}
where $d$ is distance between Tx and Rx, $d_1$ is a parameter set to 18 meters, and $d_2$ to 36 meters. Note that UMi LOS probability model distinguished between LOS and a generic non-LOS state, where LOS blockage is assumed to occur predominantly due to static objects (i.e., akin to NLOSb).
For illustration purposes, we use the states generated by the two models to calculate the path loss for a Tx-Rx pair that moves apart at 1 m/s starting from 1 to 500 meters. We use the parameters for urban medium density (Figs.~\ref{fig:medUrban} and~\ref{fig:medUrbanTrans}).
For LOS and NLOSb path loss parameters, we use the values based on measurements reported in~\cite{boban14TVT}.
For NLOSv, we use the multiple knife-edge attenuation model described in~\cite{boban14TVT}. However, to make the difference between states more clearly visible, we simplify the NLOSv model so that it adds a constant 8~dB attenuation compared to free-space path loss. Also for clarity reasons, we show path loss only (i.e., without shadow fading).
 Fig.~\ref{fig:ComparisonPL} shows the path loss results for the proposed model and 3GPP UMi LOS probability model~\cite{series2009guidelines}. Since UMi model does not model dependency on the previous LOS state, the number of transitions between the states is considerably higher than in the proposed model, particularly when the probability of LOS is close to 50\% (i.e., between 50 and 100 meters). This is an unrealistic behavior, since two vehicles can not move between LOS and non-LOS states so rapidly. The proposed model, on the other hand, takes into account the previous state and smoothly evolves the link between LOS, NLOSb, and NLOSv states. While Fig.~\ref{fig:ComparisonPL} shows a single realization of state changes and resulting path loss, we ran simulations for a large number (10$^5$) of V2V pairs with distances between 0 and 500 m. UMi model resulted in an average state change every 5 seconds, while the proposed model averaged one state change every 17 seconds. 

 The repercussion of a more realistic LOS evolution are manifold. Specifically, performing simulations without the proposed model and using simple probability models such as UMi results in:   
 i) inaccurate estimate of interference, since the calculated interference contributions from specific vehicles will be more sustained than what a simple model estimates;
 ii) overestimating the benefits of retransmissions; 
 a link in NLOS state is likely to stay in that state for a longer period of time than what is estimated by a simple model, thus making the retransmissions less effective;
 iii) erroneous estimate of performance for applications requiring continuous transmission between two vehicles, since the link duration will be impacted by the unrealistically high number of transitions between the states.

The result in  Fig.~\ref{fig:ComparisonPL} also indicates why 
shadowing correlation models 
are not sufficient for modeling time- and space- evolved V2V links. Shadowing decorrelation distance in urban and highway environment for V2V links is on the order of tens of meters (e.g., Abbas et al.~\cite{abbas2015measurement} report decorrelation distances of 5~m in urban and 30~m in highway environment) and is assumed do be independent of the Tx-Rx distance. 
However, Fig.~\ref{fig:ComparisonPL} shows that a single decorrelation distance value cannot capture the changing behavior as Tx-Rx distance changes: at low distances, the LOS decorrelation distance is high and decreases with increasing Tx-Rx distance, whereas NLOSb decorrelation distance increases with increasing Tx-Rx distance. By using distance-dependent transition probabilities, the employed model is capable of capturing this behavior.

\section{Conclusions} \label{sec:Conclusions}
We performed a comprehensive analysis of LOS blockage evolution 
for V2V links in real cities and highways. 
To efficiently model the time evolution of LOS blockage 
for V2V links, we employed a three-state Markov chain, which 
we trained 
using a large set of realistically simulated V2V links (more than 10$^6$ V2V links in high density scenarios) in urban and highway environments. 
To enable simple incorporation of LOS blockage evolution in 
simulations, we performed curve-fitting of the model parameters. The resulting LOS probability and transition probability parameters provide a detailed and realistic evolution of LOS blockage that is a function of Tx-Rx distance, environment (highway/urban), and vehicle density (low/medium/high). 

While we focused on V2V communication, the methodology we presented can be used for other types of communication (e.g., V2I, D2D with pedestrians carrying the devices, etc.), provided that realistic mobility and map information is available. 
Future work will include analysis and efficient modeling of cross-correlation 
of LOS blockage evolution for 
spatial consistency of multiple V2V links in geographic proximity.

\bibliographystyle{IEEEtran}

\end{document}